\documentclass[twocolumn,superscriptaddress,aps,showpacs,amsmath,footinbib,bibnotes]{revtex4-1}
\pdfoutput=1
\usepackage{graphicx}
\usepackage{amssymb,amsmath}
\usepackage{textcomp} 
\usepackage[bold]{hhtensor}
\usepackage{natbib}
\usepackage{lipsum}
\usepackage{bm}
\usepackage{hyperref}
\usepackage{multirow}
\usepackage{float}
\usepackage{amsmath}
\usepackage{amssymb}
\usepackage{stmaryrd}
\usepackage{mathrsfs}
\usepackage[dvipsnames]{xcolor}
\hypersetup{
	colorlinks,
	linkcolor={red!80!black},
	citecolor={Blue},
	urlcolor={MidnightBlue}
}

\begin{document}

\author{Kevin Luke}
\affiliation{HyperLight, 501 Massachusetts Avenue, Cambridge, Massachusetts 02139, USA}
\author{Prashanta Kharel}
\affiliation{HyperLight, 501 Massachusetts Avenue, Cambridge, Massachusetts 02139, USA}
\author{Christian Reimer}
\affiliation{HyperLight, 501 Massachusetts Avenue, Cambridge, Massachusetts 02139, USA}
\author{Lingyan He}
\affiliation{HyperLight, 501 Massachusetts Avenue, Cambridge, Massachusetts 02139, USA}
\author{Marko Loncar}
\affiliation{HyperLight, 501 Massachusetts Avenue, Cambridge, Massachusetts 02139, USA}
\affiliation{John A. Paulson School of Engineering and Applied Sciences, Harvard University, Cambridge, Massachusetts 02138, USA}
\author{Mian Zhang}\email{mian@hyperlightcorp.com}
\affiliation{HyperLight, 501 Massachusetts Avenue, Cambridge, Massachusetts 02139, USA}

\date{\today}
\title{Wafer-scale low-loss lithium niobate photonic integrated circuits}

\begin{abstract}
Thin-film lithium niobate (LN) photonic integrated circuits (PICs) could enable ultrahigh performance in electro-optic and nonlinear optical devices. To date, realizations have been limited to chip-scale proof-of-concepts. Here we demonstrate monolithic LN PICs fabricated on 4- and 6-inch wafers with deep ultraviolet lithography and show smooth and uniform etching, achieving 0.27 dB/cm optical propagation loss on wafer-scale. Our results show that LN PICs are fundamentally scalable and can be highly cost-effective.
\end{abstract}
\maketitle

\section{Introduction}

Thin-film lithium niobate (LN) photonic integrated circuits (PICs) have recently emerged as a promising photonics platform for many emerging applications due to their superior electro-optic performance and large second order optical nonlinearity. This is achieved through the recent development of high-confinement waveguides with low propagation loss \cite{wu_lithium_2018, liang_high-quality_2017, krasnokutska_ultra-low_2018, zhang_monolithic_2017, siew_ultra-low_2018, wolf_scattering-loss_2018}, comparable to that of passive material platforms. The desired low loss and nonlinear material properties can readily complement existing platforms such as indium phosphide (InP) and silicon (Si) photonics, where intrinsic second order nonlinearity is lacking. At device level, modulators with ultralow voltage and/or bandwidth beyond 100 GHz have been demonstrated \cite{wang_integrated_2018, he_high-performance_2019, mercante_110_2016, mercante_thin_2018, weigel_bonded_2018, rao_high-performance_2016}. Novel nonlinear optical components including frequency converters and frequency comb generators have also been realized at chip level \cite{chang_thin_2016, luo_-chip_2017, wang_ultrahigh-efficiency_2018, wang_monolithic_2019, he_self-starting_2019}. These high performance, fundamental building blocks have the potential to enable many new applications in optical communication \cite{wooten_review_2000, pfeifle_coherent_2014}, microwave photonics \cite{poberaj_lithium_2012, marpaung_integrated_2019}, quantum photonics \cite{obrien_optical_2007, alibart_quantum_2016, witmer_high-_2017, jiang_efficient_2020}, and sensing \cite{shams-ansari_integrated_2020}.

A major outstanding challenge is fabricating LN PICs at wafer-scale, i.e. if low optical loss devices can be achieved uniformly over large areas on a wafer with high throughput. Wafer-scale fabrication would enable large-scale and complex electro-optic and nonlinear optical PICs required for applications such as quantum photonics and integrated microwave photonics. In addition, a scalable process would enable a massive reduction of device cost, especially for cost-sensitive applications such as optical communications. Currently, low loss LN PIC demonstrations have only been realized for individual devices and  circuits spanning over small individual chip areas. Existing techniques employ serial device patterning techniques such as electron beam (e-beam) lithography \cite{zhang_monolithic_2017, krasnokutska_ultra-low_2018,liang_high-quality_2017,siew_ultra-low_2018} and/or complex polishing techniques \cite{wu_lithium_2018, wolf_scattering-loss_2018}. While these approaches are very effective for device prototyping, scaling these fabrication methods could require prohibitively long write times and poses major challenges for device yield. 

 \begin{figure}[]
    \centering
    \includegraphics[width=\columnwidth]{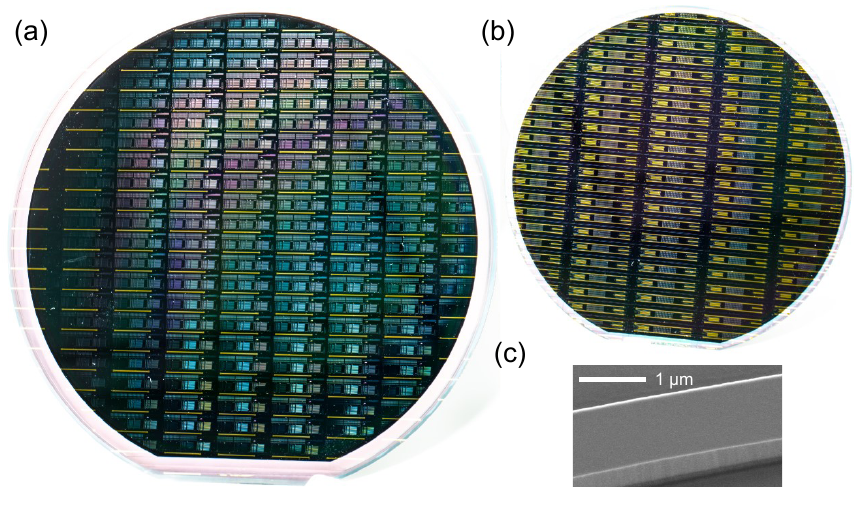}
    \caption{Photographs of 6-inch (a) and 4-inch (b) thin-film lithium niobate wafers fabricated using deep-ultraviolet lithography and standard etching processes. (c) SEM image showing typical device sidewall roughness, which is comparable to devices made with e-beam lithography \cite{zhang_monolithic_2017}.}
    \label{fig:wafer}
 \end{figure}
 
Optical lithography and direct etching for lithium niobate has been previously investigated but typically results in rough etched sidewalls \cite{ulliac_argon_2016}. Attempts to etch LN with standard fluorine (F) based etching techniques produce lithium fluoride (LiF) byproducts which are non-volatile and impede the etching process. Therefore, etching techniques for these demonstrations are typically physical etching [e.g. Argon (Ar) based]. However, interaction between Ar and photoresist could results in micromasking in the photoresist polymer, which is transferred to the etched sidewalls as roughness, especially for large etching depths. The resulting sidewall roughness increases scattering losses, which ultimately limits optical propagation loss in photoresist masked LN devices.

 \begin{figure*}[]
  \includegraphics[width=\textwidth]{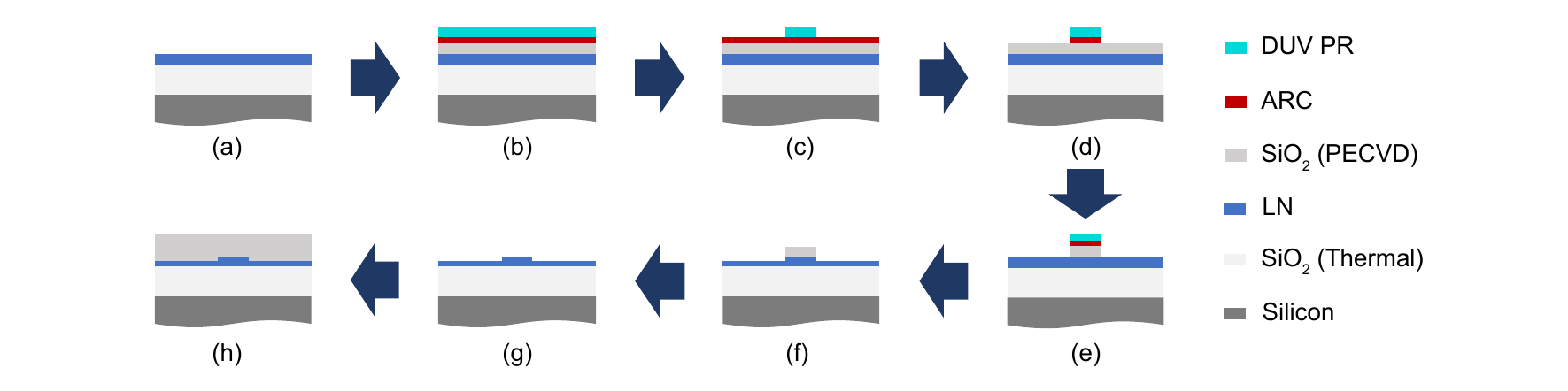}
    \caption{ Schematic of the fabrication process.  On a thin-film LN wafer (a), we deposited a $\mathrm{SiO}_2$ hard mask, then spin-coated anti-reflective coating (ARC) and DUV photoresist (b).  After DUV patterning (c) and ARC etching (d), the pattern was transferred into the $\mathrm{SiO}_2$ hard mask (e), and then into the LN layer, leaving a thin slab of LN.  The photoresist was stripped (f), and then  residual hard mask was removed (g).  Finally, the devices were cladded with $\mathrm{SiO}_2$ (h).
    }
    \label{fig:fab}
 \end{figure*}
 
\section{Device Fabrication}

Here we demonstrate deep-ultraviolet (DUV) optical lithography defined thin-film LN PICs etched on 4-inch and 6-inch wafers with propagation loss averaging 0.27 dB/cm at telecom wavelengths, approaching some of the best chip-scale demonstrations to date \cite{wu_lithium_2018, liang_high-quality_2017, krasnokutska_ultra-low_2018, zhang_monolithic_2017, siew_ultra-low_2018, wolf_scattering-loss_2018}. To address the photoresist smoothness problem in physical etching processes, we developed and employed a two-step masking technique that involved transfer of a DUV lithography defined polymer mask to a hard mask of silicon dioxide (SiO$_2$) deposited onto the LN thin-film. We transferred the polymer resist pattern to SiO$_2$ using a standard fluorine based dry etching process, and we then etched the LN layer with Ar inductively-coupled plasma (ICP) reactive-ion etching (RIE)  etching  similar to that of e-beam resist patterned devices \cite{zhang_monolithic_2017}. We then used this wafer-scale approach to fabricate a variety of optical devices, including electro-optic modulators, micro-ring resonators, and directional couplers (Fig. \ref{fig:wafer}). While the exposure time of these devices with DUV was less than a minute, exposure with an e-beam process (e.g. using an e-beam current of 10 nA on a 125 kV lithography system) would have required more than 8 days of continuous writing.

The wafer (NanoLN) consisted of a 500 $\mu$m thick silicon handle, a 4.7 $\mu$m thick thermal $\mathrm{SiO}_2$ layer, and a 600 nm thick x-cut LN thin-film (Fig. \ref{fig:fab}a). We first deposited a 650 nm layer of $\mathrm{SiO}_2$ on top of the thin-film LN via plasma enhanced chemical vapor deposition (PECVD), spun 60 nm of anti-reflective coating (ARC) and 600 nm of DUV photoresist (Fig. \ref{fig:fab}b), and then patterned the wafer with a DUV stepper of 248 nm wavelength (Fig. \ref{fig:fab}c). The ARC was etched with  standard dry etching with Ar and O$_2$ (Fig. \ref{fig:fab}d), and then the patterned DUV photoresist was transferred into the $\mathrm{SiO}_2$ hard mask using standard $\mathrm{SiO}_2$ dry etching methods in $\mathrm{C}_3\mathrm{F}_8$ chemistry (Fig. \ref{fig:fab}e). We etched LN using reactive ion etching with Ar ions \cite{zhang_monolithic_2017} and then removed the photoresist (Fig. \ref{fig:fab}f) and the $\mathrm{SiO}_2$ hard mask (Fig. \ref{fig:fab}g) using hydrofluoric acid, leaving a thin LN slab (typically 200-300 nm, depending on the desired device).  The wafer was then cladded by depositing 800 nm of $\mathrm{SiO}_2$ via PECVD (Fig. \ref{fig:fab}h).  Fig. \ref{fig:wafer}c shows the etched sidewall of a LN waveguide fabricated with this process, with sidewall roughness comparable to devices made with e-beam lithography.

\section{Measurement}

\begin{figure*}[]
  \includegraphics[width=\textwidth]{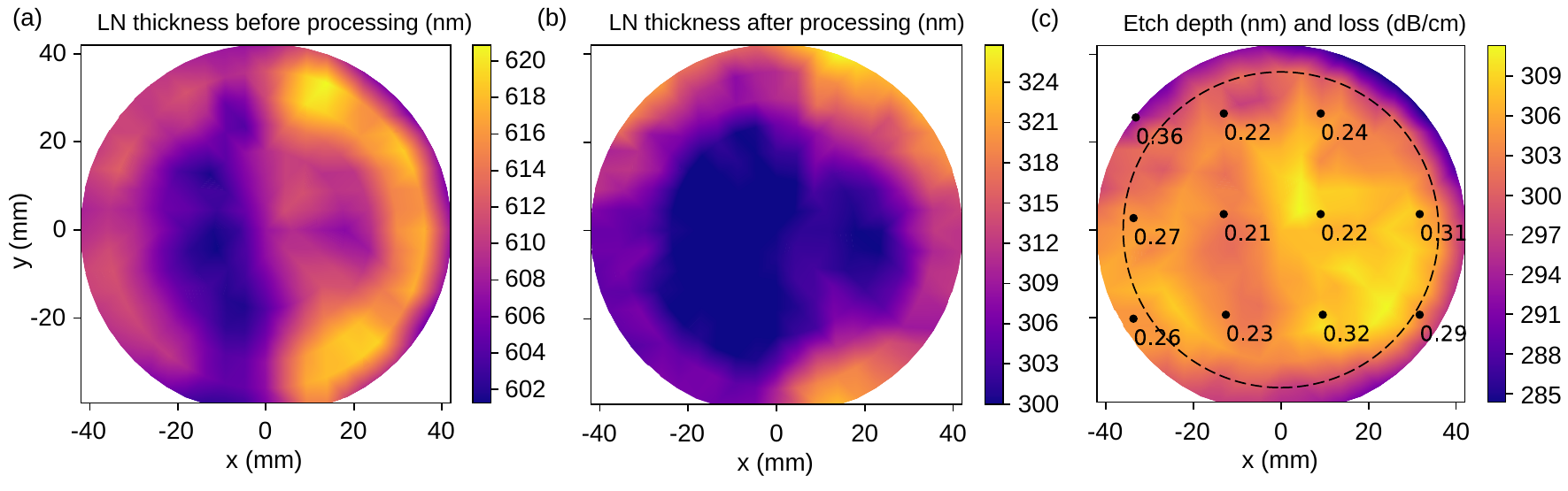}
    \caption{Measurement of LN thickness uniformity of a 4-inch wafer (a) before device processing and (b) after device processing.  (c)  Map of the etch depth, which is the difference between (a) and (b).  The etch depth is very uniform, with standard deviations of 5.9 nm across the wafer and of 3.2 nm within the dotted circle, 6 mm from the edge of the measurable wafer area.  Note that the etch depth variation is comparable to the thickness variation of the initial wafer, demonstrating that the processing was not a dominant source of nonuniformity.  Overlaid on the etch depth (c) are measured propagation loss values (in units of dB/cm) from a similarly processed wafer, showing achieved propagation losses between 0.21 dB/cm and 0.36 dB/cm, with an average of 0.27 dB/cm.}
    \label{fig:map}
 \end{figure*}

We analyzed the etched film thickness and measured a standard deviation of 5.9 nm for a 300 nm etch on a 4-inch wafer. We also show that our processing was not the dominant source of film thickness variation. We focus our discussion on a 4-inch wafer here due to our instrument limitation for characterizing 6-inch wafer sizes. We measured the LN film thickness before processing (Fig. \ref{fig:map}a), and again after etching and mask removal, before the final $\mathrm{SiO}_2$ cladding was deposited (Fig. \ref{fig:map}b). Because of the relatively large spot size of the white light interferometer used for film thickness measurement (FilMetrics F50-EXR), as well as a roughened rim due to thin-film LN wafer production process, the measurable area on the 4-inch wafer had an 8 mm edge exclusion.  From the difference of these two thickness measurements, we extracted the etch depth (Fig. \ref{fig:map}c), which shows that our processing did not introduce significant additional thickness variation.  Moreover, most of the film thickness variation after etching was located near the edge of the wafer.  Excluding 6 mm further from the edge of the measurable area (within dotted area of Fig. \ref{fig:map}c), the film thickness standard deviation was only 3.2 nm. The variation at the edge was most likely attributed to a combination of initial thickness variation and reduced chemical exposure at the edge of the wafer due to wafer handling during processing. This can be improved in the future as thin-film LN wafer production techniques advance in wafer uniformity and as wafer handling becomes automated.

\begin{figure*}[]
  \includegraphics[width=\textwidth]{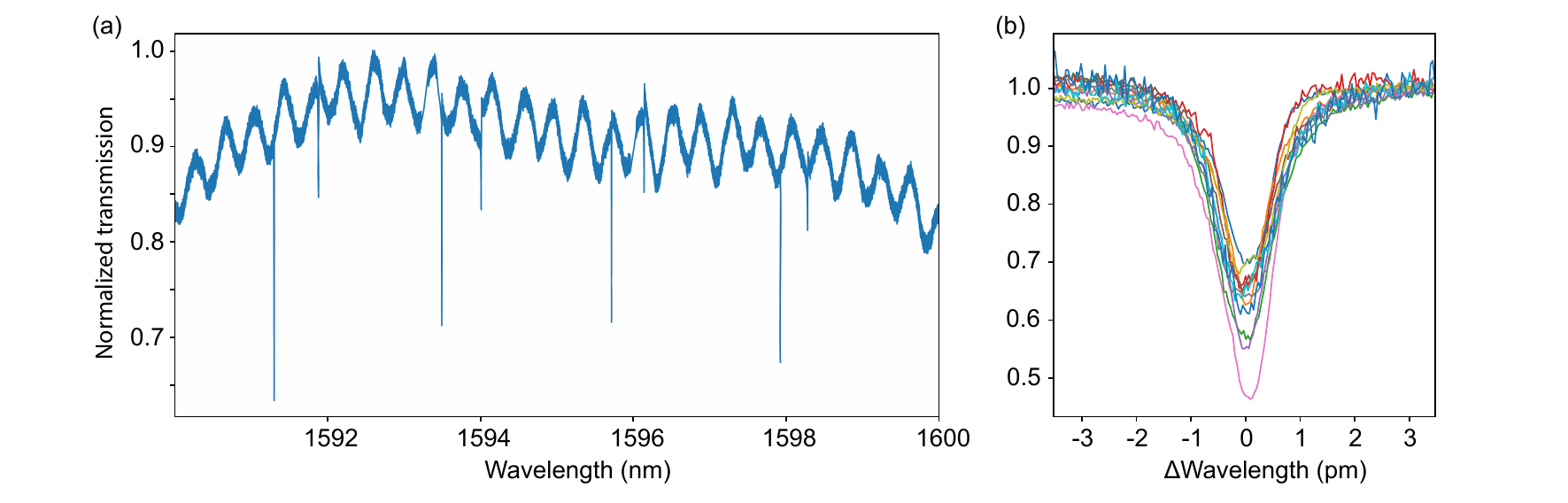}
    \caption{(a) Typical resonance spectrum of a grating-coupled micro-ring resonator.  We measure from 1590 to 1600 nm wavelength to overlap with the peak of the grating-coupler bandwidth for these devices.  (b) Micro-ring resonance spectra from different locations on the wafer (see Fig. \ref{fig:map}c for measurement locations), after renormalizing and centering around the resonant wavelength.  Each resonance is from a different reticle exposure across the wafer.  The minimum transmission varies because the waveguide-ring coupling is sensitive to fabrication, which changes the loading condition of the resonator.  However, note that the linewidths of the resonances are consistent with each other, which suggests that the fabrication is uniform across the wafer.}
    \label{fig:spectra}
 \end{figure*}

We measured an average propagation loss of 0.27 dB/cm in the etched optical waveguides across a 4-inch wafer, with a standard deviation of 0.05 dB/cm. In order to characterize the optical propagation loss, we included optical micro-ring resonators in the 22 mm by 22 mm DUV reticle that was stepped across a wafer. We coupled light from a tunable laser source into and out of the grating-coupled resonators using a vertical fiber array and measured the output power on a photodiode, obtaining the device transmission spectra for devices at various locations on the wafer (Fig. \ref{fig:spectra}). The laser wavelength range of 1590 to 1600 nm was chosen to overlap with the peak of the grating coupler bandwidth, which was designed to overlap with our laser source.  In order to avoid possible artificial linewidth narrowing due to the photorefractive effect \cite{jiang_fast_2017}, we reduced optical power (typically $< 20$ $\mu$W estimated in the device) and scanned the laser from long to short wavelength, so that the laser scan would follow the power dependent photorefractive blue shift in wavelength. Thus our linewidth measurement is a conservative upper bound estimate on the optical propagation loss. We confirmed the minimization of photorefractive effect by producing spectra with identical linewidths for both red and blue laser scan directions. Note that at these low power levels, red-shifting thermo-optic effect is not measurable. The lowest loss was measured on micro-ring resonators with etch depth of 400 nm, bending radius of 80 $\mu$m, and waveguide width of 2.0 $\mu$m near the center of the wafer, which had an intrinsic quality factor of 1.8 million, corresponding to a propagation loss of 0.21 dB/cm.  We overlaid resonance spectra from each reticle exposure on the wafer (Fig. \ref{fig:spectra}b), and they are consistent in linewidth, although the minimum transmission varies because of inherent sensitivity to resonator loading in the resonator coupling gap due to fabrication variation. We further characterized propagation loss uniformity across the full wafer (Fig. \ref{fig:map}c) and measured a maximum value of 0.36 dB/cm at the edge of the wafer, and an average of 0.27 dB/cm, with standard deviation of 0.05 dB/cm.

\section{Discussion}
At a nominal propagation loss value of $< 0.3$ dB/cm, many applications including electro-optic modulators and frequency converters can now be produced economically and at scale. There is still tremendous interest to further reduce the loss at or below what has been achieved at the single device level. The optical propagation loss achieved in this demonstration was likely limited by etching roughness \cite{ulliac_argon_2016, zhang_monolithic_2017}. The optical loss can be expected to improve further by developing a smoother hard mask transfer process, which has not been optimized in this study and has been shown to produce waveguide loss $< 1$dB/m \cite{ji_ultra-low-loss_2017}. Achieving such level of losses would enable a new library of optical components that are not currently accessible, such as near-lossless cascaded electro-optic devices and/or long on-chip optical delay lines exceeding meters of lengths. 

Our demonstration has also opened up new opportunities for high throughput wafer-scale testing capabilities that dramatically sped up the development of silicon photonics \cite{chrostowski_silicon_2015} using probes and grating couplers \cite{krasnokutska_high_2019}. This work (Fig. \ref{fig:wafer}a,\ref{fig:wafer}b) has also shown that metalization processes, as expected, are insensitive to the change on the optical waveguide layer. This enables the possibility of ultrahigh speed electro-optic devices characterized at wafer level in the near future, which is key to shortening the development cycle of LN PICs.

\section{Conclusion}

Our results show that LN PICs with low optical loss can be fabricated with good uniformity on wafer-scale with high throughput. While the optical loss and film thickness variation still have room for improvement compared to the material limit of LN and uniformity achieved on SOI respectively, our work serves as a first step to enable large-scale, complex, and low loss electro-optic and nonlinear PICs with high yield.  This can boost development in emerging large-scale PIC applications such as quantum photonics and photonic neural networks \cite{harris_linear_2018}. Moreover, the high throughput fabrication process can dramatically reduce device cost, enabling LN PIC technology to perform in a broader range of cost-sensitive applications in data- and telecommunications, sensing, and microwave photonics. Beyond monolithic LN PICs, the standard lithium niobate on insulator (LNOI) structure and the excellent passive optical performance may spur interests in achieving heterogeneously integrated optical systems on thin-film LN with laser and detector integration to achieve best-in-class performance.

\section*{Acknowledgments}
We acknowledge Dr. Fan Ye for help with measurements. This work was performed in part at the Cornell NanoScale Facility, which is supported by the National Science Foundation (Grant NNCI-1542081), and at the Center for Nanoscale Systems (CNS), which is supported by the National Science Foundation under NSF award no. 1541959. CNS is part of Harvard University. Both facilities are members of the National Nanotechnology Coordinated Infrastructure (NNCI).

\bibliography{reference}
\end{document}